\documentclass{ws-procs961x669} 
\usepackage{amssymb, amsmath}
\usepackage{mathtools}
\def\be{\begin{equation}}
\def\ee{\end{equation}}
\def\ba{\begin{eqnarray}}
\def\ea{\end{eqnarray}}
\def\nn{\nonumber}
\def\lb{\label}
\def\dfrac{\displaystyle\frac}

\def\E{{\cal E}}
\def\e{{\rm e}}
\def\a{\kappa}
\def\z{\zeta}
\def\k{{\sigma}}

\renewcommand{\theequation}{\arabic{section}.\arabic{equation}}
\begin{document}
\title{  
 Einstein-Maxwell-Dilaton-Axion mass formulas for black holes\\ with struts and strings}
\author{Dmitri Gal'tsov$^{a}$, G\'erard Cl\'ement$^b$, Igor Bogush$^a$  }
 
\address{$^a$ Faculty of Physics, Moscow State University, 119899, Moscow, Russia \\$^b$ LAPTh, Universit\'e Savoie Mont Blanc, CNRS, France\\
Email: galtsov@phys.msu.ru,  gclement@lapth.cnrs.fr,  igbogush@gmail.com}

\begin{abstract}
Mass formulas are obtained for stationary axisymmetric solutions of the Einstein-Maxwell dilaton-axion theory, which have a regular rod structure on the axis of symmetry. Asymptotic mass, angular momentum and charge are expressed as the sums of masses, angular momenta and charges of rods dressed with field contributions. The calculation is based on a three-dimensional sigma model representation of the stationary EMDA system and the Tomimatsu approach proposed for the Einstein-Maxwell system. Our results provide an alternative interpretation of mass formulas and thermodynamics for black holes with Dirac and Misner strings. It is also applicable to aligned multiple black holes with struts.  
\end{abstract}
 

\bodymatter

\section{Introduction}
 
Mass formulas for black holes and the concept of irreversible mass were proposed  by Christodoulou\cite{Christodoulou:1970wf}, Hawking\cite{Hawking:1971vc}, Christodoulou and Ruffini\cite{Christodoulou:1971pcn},  and Smarr\cite{ Smarr:1972kt} in the early 1970s, shortly before Hawking discovered the evaporation of a black hole \cite{Hawking:1974rv}. They have played an important role in understanding the energy extraction from rotating black holes and the energy balance of merging black holes, which was brilliantly confirmed in the recent  experiments of Ligo. 
The thermodynamics of black holes\cite{Bekenstein:1972tm,Bardeen:1973gs,Hawking:1976de,Gibbons:1976ue} then gave them a deep quantum interpretation. 
The mathematical foundations and detailed derivation of mass formulas in the Einstein-Maxwell theory were given by Carter \cite{Carter:1973rla}.
 
The original integral Smarr's mass formula\cite{Smarr:1972kt} relates the total values of  mass,  angular momentum and
electric charge of  black holes in Einstein-Maxwell theory with the  horizon area. The area term in this formula was originally interpreted  mechanically as the work of {\em stresses} of the horizon. Further, this interpretation was forgotten in favor of the thermodynamic  one. When interest turned to solutions with NUT, it was immediately found
\cite{Hunter:1998qe,Hawking:1998jf,Carlip:1999cy,Mann:1999pc}, 
that the Misner string
also contributes to the Smarr  mass formula, and this contribution was included in the entropy term. This interpretation was revived recently   in  slightly different terms in a series of papers\cite{Hennigar:2019ive,Durka:2019ajz,Wu:2019pzr,Chen:2019uhp, BallonBordo:2020mcs}, based on the Bonnor   interpretation of the Misner strings as mild physical singularities \cite{Clement:2015cxa}.

In our opinion, the most natural description of solutions with Misner  and
Dirac strings as well as aligned multicenter solutions with struts   can be given in terms of the rod structure, introduced by Harmark\cite{Harmark:2004rm}, developing earlier ideas of Emparan and Reall\cite{Emparan:2001wk}.
The main novelty of this approach is the derivation of separate formulas for the partial masses for each component of the entire system, represented by a set of rods. In this interpretation, Misner's strings appear as independent components on the same base and as black holes. 

Using the representation of vacuum stationary axisymmetric solutions  in Weyl coordinates, we  consider the solutions as generated by data on the polar axis, which look like distributional matter objects that are sources of the Poisson equation for the gravitational potential\cite{Israel:1976vc}. Accordingly, the total gravitational field can be viewed as a nonlinear superposition of the components such as black holes, struts, Misner and Dirac strings. 
Despite the nonlinearity of Einstein's equations, the total mass, angular momentum (and electric charge in the case of an electrovacuum) can be represented as the sum of the individual contributions of the constituents. This simple additivity is associated with conservation laws for the Komar and Gauss integrals. 
The magnetic charge and magnetic mass (NUT) in this description are due to the Misner and Dirac strings, which correspond to the individual elements in the rod system, so these charges do not enter the black hole horizon mass. Their contribution enters the asymptotic Komar mass as the proper masses  of the Dirac and Misner strings along with the black hole contribution.  
 
The difference between the horizon  and the string rods is that the directions of the former are timelike, and the directions of the latter are spacelike. Both rods are Killing horizons and have an associated surface gravity. But the entropic interpretation of the surface contribution of spacelike rods (often encountered) does not seem convincing. An alternative interpretation may be similar to Smarr's original point of view. 
 
This programme was previously performed for the Einstein-Maxwell system\cite{Clement:2017otx,Clement:2019ghi}, and here it is extended to Einstein-Maxwell-dilaton-axion gravity (EMDA). We show that Tomimatsu's proposal within the context of  electrovacuum  \cite{Tomimatsu:1983qc, Tomimatsu:1984pw} on calculating Komar integrals over rods in terms of the boundary values of Ernst's potentials can be generalized to the sigma model representation of stationary dilaton-axion gravity. Surprisingly, the obtained mass formulas are very similar to the formulas obtained in the case of Einstein-Maxwell. 
 
\section{Stationary EMDA gravity}
The EMDA gravity can be viewed as a consistent truncation of a toroidal reduction of the heterotic string  \cite{Sen:1992ua}, or as a truncation of $N=4$ four-dimensional supergravity \cite{Gibbons:1982ih}\cite{Kallosh:1994ba}.  Rotating electrically charged black hole solution of EMDA theory was obtained by Sen  \cite{Sen:1992ua} within the first approach and  independently, with the inclusion of NUT and magnetic charge, by Gal'tsov and Kechkin \cite{Galtsov:1994pd} in the second context.  Recently, these solutions have attracted attention as an alternative to the Kerr metric in an effort to find astrophysical evidence for new physics  \cite{Narang:2020bgo,Banerjee:2020qmi,Banerjee:2020ubc}. 

The mass formulas and thermodynamics of EMDA black holes have been discussed frequently in the past, though not in full generality,  see, for example  \cite{Rogatko:1998kj,Ghosh:2008xc} (a more complete list of references prior to 2008 can be found in \cite{Ghosh:2008xc}). Our discussion here, based on generalization of~ \cite{Clement:2019ghi},  covers solutions with Misner and Dirac strings, and is also applicable to hitherto unknown solutions for double black holes with struts that are expected to exist.
  
 The EMDA action in our conventions reads:
\begin{equation}\label{ac1}
S=\frac{1}{16\pi}\int \left\{R-2\partial_\mu\phi\partial^\mu\phi -
\frac{1}{2} \e^{4\phi}
{\partial_\mu}\a\partial^\mu \a
-\e^{-2\phi}F_{\mu\nu}F^{\mu\nu}-\a F_{\mu\nu}{\tilde F}^{\mu\nu}\right\}
\sqrt{-g}d^4x,
\end{equation}
where ${\tilde F}^{\mu\nu}=\frac{1}{2}E^{\mu\nu\lambda\tau}F_{\lambda\tau}$.
Dilaton $\phi$ and axion $\kappa$   parameterize a coset
 $SL(2,R)/SO(2)$, and the $SL(2,R)$ group  is the symmetry of the full action. To see this, one can introduce the complex scalar 
\begin{equation}\lb{zdef}
\z = \a + i \e^{-2\phi}
\end{equation}
and the self-dual Maxwell field ${\cal F}=\left(F+i{\tilde F}\right)/2$. Then the action takes the form
\begin{equation}
S=\frac{1}{16\pi}\int \left\{R-2\big|\partial \z (\z-{\bar \z})^{-1}\big|^2+\left(i \z {\cal F}_{\mu\nu}{\cal F}^{\mu\nu}+c.c.\right)  \right\}
\sqrt{-g}d^4x,
\end{equation}
which is invariant under the $SL(2,R)$ transformations
\begin{align}
    &\z\to \frac{\alpha \z+\beta}{\gamma \z+\delta},\qquad \alpha\beta-\gamma\delta=1,\nonumber\\
    & F\to (\gamma \a +\delta)F+\gamma\e^{-2\phi}\tilde{F},
\end{align}
which interchange the modified Maxwell equations and the Bianchi identity 
\begin{align}\lb{maxw}
&\nabla_\nu G^{\mu\nu} =0,\quad \nabla_\nu  {\tilde F}^{\mu\nu}=0,   \\&  G^{\mu\nu}=  e^{-2\phi}F^{\mu\nu}+\a{\tilde F}^{\mu\nu}.
\end{align}
 
The Einstein equations are
\begin{equation}\lb{Ric}
R_{\mu\nu}=2\phi_{,\mu}\phi_{,\nu}+
\frac{1}{2}\e^{4\phi}\a_{,\mu}\a_{,\nu}-
\e^{-2\phi}\left(2F_{\mu\lambda}{F^\lambda}_\nu +\frac{1}{2}F^2g_{\mu\nu}\right).
\end{equation}
For stationary and axisymmetric configurations  the axion and dilaton kinetic term disappear from the Einstein equations in the  $t-\varphi$ sector, and the dilaton enters only through the scale factor $\e^{-2\phi}$ in front of the Maxwell energy-momentum tensor.  This is crucial for our derivation.  
 
Assuming the existence of the Killing vector $k=\partial_t$, the metric can be represented as 
\begin{equation}
ds^2=g_{\mu\nu}dx^\mu dx^\nu=-f(dt-\omega_idx^i)^2+\frac{1}{f}h_{ij}dx^idx^j,
\end{equation}
where three-dimensional metric $h_{ij}$, three-dimensional rotation  vector $\omega_i,\; (i, j=1, 2, 3)$ and the scale factor $f$ depend only on the space coordinates $x^i$.
The spatial parts of the Bianchi identities and the modified Maxwell equations   can be solved by introducing electric $v$  and magnetic $u$ potentials
\begin{align}\lb{vdef}
&F_{i0}=\frac{1}{\sqrt{2}}\partial_iv,\\\lb{udef}
&G^{ij}=-\frac{f}{\sqrt{2h}}\epsilon^{ijk}
\partial_ku.
\end{align}
The mixed $R_0^i$   Einstein equations are solved by introducing the twist potential $\chi $ obtained by dualizing  the rotation two-form:
\begin{equation}\lb{eq:chi}
f^2h_{il}\frac{\epsilon^{ljk}}{\sqrt{h}}\partial_j\omega_k,=\partial_i\chi +v\partial_iu-u\partial_iv.
\end{equation}
The remaining equations reduce to those of three-dimensional gravitating sigma-model equations for six scalars $ X^A=f,\chi,v,u,\phi,\a, \;A=1,\ldots,6,$ and the three-dimensional metric $h_{ij}$:  
\begin{equation}
S=\int \left[{\cal R}(h)-{\cal G}_{AB}(X)\,\partial_i X^A\,\partial_j X^B
\,h^{ij}\right]\sqrt{h}\,d^3x,
\end{equation}
where the target space metric ${\cal G}_{AB}$ can be presented as
\begin{eqnarray}
{dl}^2 & = & \frac{1}{2}f^{-2}[{df}^2+
(d\chi +vdu-udv)^2]-f^{-1}[e^{2\phi }{(du- \a dv)}^2
+e^{-2\phi }{dv}^2] \nonumber \\
& + & 2{d\phi }^2+
\frac{1}{2}e^{4\phi }{d \a }^2.
\end{eqnarray}
The isometry group of this metric is $SO(3,2)$, as was identified in \cite{Galtsov:1994sjr}. Based on the isomorphism $SO(3,2)\sim Sp(4,R)$, a convenient $4\times 4$ matrix representation of the   coset $Sp(4,R)/SO(1,2)$ was suggested \cite{Galtsov:1995tnt}, suitable for the generation technique.
In terms of complex coordinates, one of which is (\ref{zdef}) and two other are the following generalization of the Ernst potentials \cite{Galtsov:1997jrl}
\be \Psi=u-\z v,\quad \E=if-\chi+v\Psi,
\ee
the target space is a three-dimensional K\"ahler  space.

\section{Rod suructure}
Consider the spacetime metric, admitting two commuting Killing vectors $k,\, m$ corresponding to stationarity and axial symmetry. Let $t,\,\varphi$ be chosen so that  $k=\partial_t,\, m=\partial_\phi$ with $t\in R$ and $\varphi\in [0, \pi]$. The remaining coordinates will be primarily assumed of the Boyer-Lindquist type, $r, \theta$ with $0<r<\infty$ and $\theta \in [0, \pi]$.
We assume that the spacetime manifold has no strong naked curvature singularities, but can have visible line singularities along (a part of) the polar axis in the sense of Israel \cite{Israel:1976vc}. These include cosmic strings (conical singularities), struts in aligned multiple black hole solutions (conical singularities of both positive or negative tension) and Misner strings in spacetimes with NUTs, without time periodicity being imposed. We also assume that spacetime is asymtotically flat, or asymptotically locally flat.  To such a spacetime one can ascribe a {\em rod structure} following Harmark \cite{Harmark:2004rm}.

Let  $\gamma _{ab},\;x^a=t,\,\varphi$ is the two-dimensional Lorentzian metric of the subspace spanned by the Killing vectors.  
Introduce  the Weyl cordinates $\rho,\, z $  such that $\rho$ is
\begin{equation}             \label{rhodef}
 \rho = \sqrt{|\det \gamma |}\,,
\end{equation}
and  $z$ to ensure the metric form:
 \begin{equation}               \label{Gzr}
 ds^2 = \gamma _{ab}(\rho, z) dx^a dx^b + e^{2\nu}(d\rho^2 +  dz^2)\,,
\end{equation}
where   $\nu$ is a function  of $(\rho, z)$. 
To find the rod structure, one has to solve the equation 
\be  \lb{detG}
\rho(r,\theta)=0.
\ee
Generically, the solution splits the polar axis into a certain number of finite or semi-infinite intervals
$(-\infty, z_1], [z_1, z_2], \dots, [z_N, +\infty)$ called rods (we will
label two semi-infinite rods by $n=\pm$, and the remaining finite ones by an index $n$ corresponding to the
left bound of the interval). Each rod can be equipped with a two-dimensional vector, called {\em rod direction}, via the following reasoning. At $\rho=0$ the matrix $\gamma_{ab}(0,z)$  is degenerate  by virtue of the definition (\ref{rhodef}), so it must have zero eigenvalues. For quasiregular spacetimes, such that on the symmetry axis there are no strong curvature singularities,   these eigenvalues must be non-degenerate except, perhaps,  for a discrete set of ``turning'' points $z_n$  which mark the ends of rods.
The eigenvector $v_n^a$, satisfying the equation 
 \begin{equation}
 \gamma _{ab}(0, z)v_n^b = 0  \,
\end{equation}
on any segment $z\in [z_n, z_{n+1}]$, is called the {\rm $n$-th rod direction}. 
By continuity, this vector can be extended to small $\rho\neq 0$, and a more accurate analysis shows 
that, in the leading order in $\rho$, its 2D norm behaves as
\be \lb{axe}
v_{n}^2=\gamma _{ab}v_{n}^a v_{n}^b\sim \pm a(z) \rho^2,\qquad  e^{2\nu}\sim c^2 a(z),\ee
where $c$ is some constant and the sign $\pm$ corresponds to {\em spacelike} and {\em timelike} rod  respectively.  

Normalization of directional vectors can be chosen in different ways. One possibility,  {\em Killing} normalization, relates to preferred normalisation an associated Killing vector in spacetime: $V=v^0k+v^1 m$, namely, $v_0=1$ in the timelike case, and $v_1=1$ in spacelike. In view of behavior near the axis (\ref{axe}), in both cases the rod itself (i.e. the submanifold $\rho=0$) will be the Killing horizon for so defined Killing vector. 
Another possibility is to choose the scale factor in such a way as to ensure the finiteness of its norm at $\rho\to 0$:
\be \lb{axe1}
l_n^a=\lim_{\rho\to 0}\rho^{-1} e^{-\nu}v_n^a,
\ee
again with an option of further constant rescaling. An important property  of so defined rod direction is its constancy (in view of (\ref{axe})) along the rod, i.e., 
\be
 \partial_z l^a_n =0.
\ee 

Timelike rods of finite length correspond to black hole horizons,  infinite timelike rods describe acceleration horizons.
The constant components of such a horizon rod are connected with  its angular velocity $\Omega_H$ and the surface gravity $\varkappa_H$ of the Killing vector 
\be
\xi =k+\Omega_H\, m\,.
\ee
 The  surface gravity \be
\varkappa_H=\left(-\xi_{\mu ;\nu} \xi^{\mu ;\nu}/2\right)^{1/2}\ee
can be expressed in the Weyl coordinates as
\be\lb{karod}
\varkappa_H= \lim_{\rho\to 0} \left(- \rho^{-2}e^{-2\nu}\gamma_{ab}v^a v^b  \right)^{1/2},
\ee
with $v^0=1,\,v^1=\Omega_H$. It may be convenient to choose a canonical normalization of the  directional vectors that corresponds to the unit surface gravity: 
\be\label{lh}
l_H=\left( 1/\varkappa_H,\;  \Omega_H/\varkappa_H\right).
\ee

Spacelike rods correspond to line defects, such as cosmic strings, struts,  Misner strings, which can be also Dirac strings associated with the corresponding vector  potentials. These rods are Killing horizons for some spacelike Killing vectors.
For them one also has  constant angular velocities $\Omega_n$ and  
{\em spacelike} surface gravities  
\be\lb{kadef}
\varkappa_n= \lim_{\rho\to 0} \left( \rho^{-2}e^{-2\nu}\gamma_{ab}v_n^a v_n^b  \right)^{1/2},
\ee
so that the normalized directional vectors will be 
\be\label{ln}
l_n=\left( 1/\varkappa_n,\;  \Omega_n/ \varkappa_n\right).
\ee
Here $v^1_n=1$.  
With the normalized spacelike directional vector, the period (\ref{ident}) will be $2\pi$. If the coordinate $\eta$ associated with the spacelike rod Killing vector   $l=\partial_\eta$, conical singularity is absent with perdiocity of $\eta$  with
\be\label{ident}
\Delta \eta=2\pi\lim_{\rho\to 0} \left(\rho^{2}e^{2\nu}\left(\gamma_{ab}l_{n}^a l_{n}^b  \right)^{-1}\right)^{1/2}.
\ee
This is important in the situation when two spacelike rods meet at the turning point, where potentially a mismatch of periodicities may lead to {\em orbifold} singularities. This may happen for instantons \cite{Chen:2010zu}.
 
 
\section{Komar charges  }\label{sec:komar}
Following  \cite{Clement:2019ghi}, we start with the Komar definition \cite{Komar:1958wp} of the asymptotic mass and angular momentum:
  \be
M_\infty = \frac1{4\pi}\oint_{\Sigma_\infty}D^\nu k^{\mu}d\Sigma_{\mu\nu}, \quad
J_\infty = -\frac1{8\pi}\oint_{\Sigma_\infty}D^\nu m^{\mu}d\Sigma_{\mu\nu}.
\lb{koMJ}
 \ee
 We also need a conserved electric charge as the surface integral. The definition of a conserved electric charge in EMDA  follows from the modified Maxwell equations:
\be  \lb{Qu}
 Q_\infty=   \frac1{4\pi}\oint_{\Sigma_\infty} G^{\mu\nu}   d\Sigma_{\mu\nu}.
\ee
Consider some stationary axisymmetric solution with an arbitrary rod structure $z_n,\, l_n$. Each rod must be surrounded by a thin cylinder $\Sigma_n$. Using Ostrogradski formula one can transform the asymptotic total mass to the sum of the local Komar masses $M_n^d$ of rods (``direct'' masses) and the bulk contribution coming from the fields:  $M = \sum_n M_n^d+M_F,$ where 

 \be
 M_n^d=\dfrac1{4\pi}\oint_{\Sigma_n} D^\nu k^\mu d\Sigma_{\mu\nu}
,\qquad M_F= \frac1{4\pi}\int D_\nu D^\nu k^\mu dS_\mu, \lb{koM1}
 \ee
where $\Sigma_n$ are the spacelike surfaces bounding the various
sources\footnote{We use the metric signature (-+++) and the
convention $d\Sigma_{\mu\nu}=1/2
\sqrt{|g|}\,\epsilon_{\mu\nu\lambda\tau} dx^\lambda dx^\tau$ with
$\epsilon_{t\rho z\varphi}=1$ in Weyl coordinates. We will label $t,
\varphi$ by an index $a$, and the remaining coordinates $\rho, z$ by
$i,j$.  The two-dimensional Levi-Civita symbol
$\epsilon_{ij}$ is defined with $\epsilon_{\rho z}=1$ and
$\epsilon_{xy}=1$ respectively.}, and the second integral is over the bulk. 
Similarly, the asymptotic angular momentum will read
$J = \sum_n J_n^d+J_F$, where
\be
 J_n^d=- \dfrac1{8\pi}\oint_{\Sigma_n} D^\nu m^\mu d\Sigma_{\mu\nu}
,\qquad J_F=- \frac1{8\pi}\int D_\nu D^\nu m^\mu dS_\mu. \lb{koJ1}
 \ee
Using  the well-known Killing lemma for $k$,
 \be
D_\nu D^\nu k^\mu = -[D_\nu,D^\mu]k^\nu = - {R^\mu}_\nu k^\nu,  
 \ee
 and similarly for $m$, and an explicit form for the Ricci tensor in EMDA theory (\ref{Ric}),
one can 
express the bulk integrals as 
\begin{align}\lb{MF}
&M_F  
=-
\frac1{4\pi}\int \e^{-2\phi}\left(F_{it}F^{it}-F_{i\varphi}F^{i\varphi}\right)\sqrt{|g|}d^3x,\\\lb{JF}
&J_F = \frac1{4\pi}\int \e^{-2\phi}F_{i\varphi}F^{it}\sqrt{|g|} \,d^3x.
\end{align}

Using the identities
\be
F_{it}{\tilde F}^{it}-F_{i\varphi}{\tilde F}^{i\varphi}=0,\qquad
F_{i\varphi}{\tilde F}^{it}=0
\ee
one can rewrite (\ref{MF}) and (\ref{JF}) as
\begin{align}\lb{MG}
&M_F  
=-
\frac1{4\pi}\int  \left(F_{it}G^{it}-F_{i\varphi}G^{i\varphi}\right)\sqrt{|g|}d^3x,\\\lb{JG}
&J_F = \frac1{4\pi}\int  F_{i\varphi}G^{it}\sqrt{|g|} \,d^3x.
\end{align}
In view of the modified Maxwell equations, these expressions can be presented in the full divergence form
\begin{align}
    &M_F =
    -\frac{1}{4\pi}\int 
        \partial_i\left[
            \sqrt{-g}
            \left(
                A_t G^{it} - A_\varphi G^{i\varphi}
            \right)
        \right]
    d^3x,\\
    &J_F =
    \frac{1}{4\pi}\int
    \partial_i\left(
        \sqrt{|g|} A_\varphi G^{it}
    \right)
    \,d^3x,
\end{align}
suggesting the representation of the asymptotic mass and angular momentum as the surface integrals over the rods \footnote{We assume that the integrals over in infinite sphere vanish, otherwise they must be added too. This happens if the North and South Misner and Dirac strings are arranged non-symmetrically, by adding suitable constants to the asymptotic values of the rotation function $\omega$ and the azimuthal component of the four-potential $A_\varphi$.}:
\begin{align}
   & M_\infty = \sum_{n} M_n,\qquad
    M_n = \frac1{8\pi}\oint_{\Sigma_n}\,
    \left(
        g^{ij}g^{ta}\partial_jg_{ta}
        + 
        2 \left(
            A_t G^{it} - A_\varphi G^{i\varphi}
        \right)
    \right)d\Sigma_i
 \\
  &  J_\infty = \sum_{n} J_n,\qquad
    J_n = -\frac1{16\pi}\oint_{\Sigma_n}\,\left(
        g^{ij}g^{ta} \partial_jg_{\varphi a}
        + 4 A_\varphi G^{it}
    \right)
    d\Sigma_i,
\end{align}
where we have also rewritten the direct Komar integrals in an explicit form. Thus we have succeeded in presenting the EMDA  bulk contributions in the form of the integrals over the rods in the same way as it was done in \cite{Clement:2019ghi} for the Einstein-Maxwell system.
This procedure can be regarded as rod's dressing by the bulk field. So the resulting Smarr formulas for the asymptotic mass and angular momentum were written as  sums of the dressed rod's contributions. Note, that in the general case, this is the only representation for the global $M,\,J$ that can be found, since different   forms of rod's contribution do not allow one to write asymptotic quantities directly in terms of physical charges as was possible for a single black hole.  

From now on we assume the standard Weyl-Papaperou parametrization of the metric
 \be\lb{weyl}
ds^2 = -f(dt-\omega d\varphi)^2 + f^{-1}[e^{2k}(d\rho^2+dz^2)+ 
\rho^2d\varphi^2],
\ee
and represent the four-potential as 
\be
\sqrt{2}\,A_\mu dx^\mu = vdt + A d\varphi
 \ee
(note that $v, u, A$ in this paper differs from these in \cite{Clement:2019ghi} by $\sqrt{2}$ to be consistent with the EMDA literature).
Computation of the direct rod Komar integrals was given in \cite{Smarr:1972kt}. It amounts to dualizing the rotation three-form  \be\lb{twist}
\partial_i\chi = -f^2\rho^{-1}\epsilon_{ij}\partial_j\omega
+ (u\partial_i v - v\partial_i u)\,,
\end{equation}
and leads to
  
\be\lb{Md}
M_n^d= \frac1{8\pi}\int_{n}\omega\left[\partial_z\chi +
(v\partial_z u
- u\partial_z v)\right]dzd\varphi
\ee
Similarly for the angular momentum
\be
J_n^d= -\frac1{16\pi}\int_{n}\omega\left\{2-\omega\left[\partial_z\chi +
 (v\partial_z u - u\partial_z v)\right]\right\}dzd\varphi.
\ee
Note the appearance of the electromagnetic potentials in the expressions for direct rod mass and angular momentum due to use of the dualized twist potential, i.e., implicit use of Einstein equations. These terms will be exactly canceled by the field contributions to the same quantities.  

 To continue, we show that,
apart from constancy (i.e. $z$-independence) of the rod angular velocity $\Omega=1/\omega$ and the generalized surface gravity $\varkappa$,   one can prove for any rod the following relations:
\begin{align} 
&\lim_{\rho\to 0} \left(G^{\rho t}-\omega G^{\rho \varphi} \right)\sqrt{|g|} =0, \lb{tfi}\\
&\lim_{\rho\to 0} \left (v+A/\omega\right)\equiv -\sqrt{2}\Phi=\rm{const,}\lb{Ph}
\end{align} 
where $\Phi$ can be interpreted as an electric potential of the rod. Together with constancy of $\omega$, the last relation gives also
\be \lb{Av}
\partial_z A=  -\omega \partial_z v,
\ee
as $\rho\to 0.$

Using (\ref{tfi}-\ref{Av}), and the EMDA dualization equation (\ref{udef})
one can rewrite the field contribution to the dressed rod mass   (\ref{MF}) as  
\be\lb{MnF}
  M_n^F=-\frac1{8\pi}\oint_{\Sigma_n}\left[\omega\left(v\partial_z u - u
\partial_z v\right) - \partial_z(uA) \right]dzd\varphi.
\ee
 Combining this with the direct Komar mass (\ref{Md}) we obtain the dressed rod's mass 
  \be\lb{Mrn}
M_n = \frac1{8\pi}\int_{\Sigma_n}\left[\omega\partial_z \chi +
 \partial_z(A\,u)\right]dzd\varphi.
 \ee
It may seem surprising that this formula does not differ from that for Einstein-Maxwell system. No explicit scalar terms are seen. 
 
For the bulk contribution to the angular momentum, one also applies the Eq. (\ref{tfi}) and then magnetic dualization equation (\ref{udef}). The subsequent transformations are the same as in \cite{Clement:2019ghi} and lead to the same result
\be
J_n^F=
 \frac1{8\pi}\int_H\left[
      \omega^2 u\partial_z v
    + \omega \partial_z(A  u)
\right]dzd\varphi.
\ee
Combining with the direct Komar momentum, we obtain for the dressed rod:
\be\lb{Jrn}
  J_n = \frac1{16\pi}\int_H\omega\left[-2 + \omega\partial_z\chi +  \partial_z(A \,u) -  \sqrt{2} \omega \Phi_n\partial_z u\right]dzd\varphi,
 \ee
this expression also does not contain the dilaton and axion. 
Now consider Komar electric charge of a rod (\ref{Qu}). Explicitly, it is given by the flux
 \be\lb{Qun}
Q_n = \frac1{4\pi}\int_{\Sigma_n}\sqrt{|g|}G^{t\rho}dzd\varphi.
 \ee
 Manipulating with indices and using dualization equation (\ref{udef})
  \be\lb{indi}
G^{t\rho} = \frac{g^{\rho\rho}}{g_{tt}}G_{t\rho} -
\frac{g_{t\varphi}}{g_{tt}}G^{\varphi i} = e^{-2k}\left[G_{t\rho}
+ \frac{f\omega}\rho \partial_z u\right],
 \ee
one finds that in the limit $\rho\to 0$ only the second term contributes, so
\be\lb{Qn}
Q_n=\frac1{4\sqrt{2}\pi}\int_{\Sigma_n} \omega\partial_z  u\,dzd\varphi.
 \ee
After integration over  $\varphi$
 \begin{align}\lb{Mrn1}
M_n =  
&  \frac{\omega_n}{4}\chi\Big {\vert}^{z_{n+1}}_{z_n} +\frac14 (A u) \Big{\vert}^{z_{n+1}}_{z_n}.
 \end{align}
Similarly,  for the angular momentum (\ref{Jrn}) one finds
\begin{align}\lb{Jrn1}
  J_n = 
 \frac{\omega_n}{8}  \left\{ -2(z_{n+1}-z_n)    + \left[\omega_n \left( \chi  -\sqrt{2}\Phi_n  u\right)+
 A u \right] \Big {\vert}^{z_{n+1}}_{z_n}\right\}.
 \end{align}
 Finally, the electric charge will be
  \be\lb{Qrn}
Q_n= \frac{\omega_n}{2\sqrt{2}} u \Big {\vert}^{z_{n+1}}_{z_n}.
 \ee
 The angular velocity of a rod is defined as a limit
 \be
 \Omega_n=\lim_{\rho\to 0}\ \omega^{-1}(\rho,z),\qquad  z_n <z<z_{n+1}.
 \ee
Combining the above formulas we obtain
rod Smarr mass
\be\label{leSm}
M_n=\frac{L_n}{2}+2\Omega_n J_n+\Phi_n Q_n,
\ee
where $L_n=z_{n+1}-z_n$ is the $n$-s rod length (for infinite rods some length regularization is needed). Please note that this formula is identical; it is valid for any rod, regardless of the specific parameter values.

\section{Length versus entropy}
Although it looks like a one-dimensional object, any rod with finite surface gravity has a finite surrounding cylindrical area: 
\be\lb{Arn}
{\cal{A}}_n=\oint d\varphi\int_{z_n}^{z_{n+1}} \sqrt{|g_{zz}g_{\varphi\varphi}|} dz =2\pi \int_{z_n}^{z_{n+1}}\sqrt{|e^{2k|}}|\omega| dz.
\ee
The surface gravity (by definition, positive) in Weyl coordinates reads
\be\label{kappa_weyl}
\kappa_n=|\Omega_n|\sqrt{|e^{-2k}|}. 
\ee
Thus the integral (\ref{Arn}) is  simply proportional to the rod's length:
\be\lb{kapA}
 \frac{\kappa_n}{2\pi}{\cal{A}}_n=  {z_{n+1}-z_n}=L_n .
\ee
This is true for both timelike rods (horizons) and spacelike ones (defects). In the first case, the surface gravity is proportional to the Hawking temperature, and the area of the horizon is proportional to the entropy
\be
T_H = \kappa_H/2\pi,\qquad S_H={\cal{A}}_H/4,
\ee
therefore
\be\lb{TS}
T_HS_H = \frac{\kappa_H}{8\pi}{\cal{A}}_H= \frac{L_H}4,
\ee
and the  Smarr formula can be rewritten
as \cite{Smarr:1972kt}
 \be\lb{smarr}
M_H = 2\Omega_H J_H + 2T_H S_H + \Phi_H Q_H.
 \ee
For spacelike rods  such an identification does not seem justified, so we prefer to leave Smarr formula in its ``length'' form (\ref{leSm}).

For multicenter solutions this mass relation holds separately for each black hole constituent. 
As stated in \cite{Clement:2017otx,Clement:2019ghi},  this is also true when black holes have magnetic and/or NUT charges. In our interpretation, the contribution of these parameters to the total asymptotic mass comes from individual rods representing the Dirac and Misner strings.
 
\section{ EMDA rotating dyon with NUT}
The metric of  the EMDA dyon with the NUT parameter constructed in 1994  \cite{Galtsov:1994pd} can be represented in the Kerr form:
\be ds^2 =
    - \frac{\Delta - a^2 \sin^2\theta}{\Sigma} \left(dt - \omega d\varphi\right)^2
    + \Sigma \left(
        \frac{dr^2}{\Delta} + d\theta^2 + \frac{\Delta \sin^2\theta}{\Delta - a^2 \sin^2\theta} d\varphi^2
    \right),\ee with the modified coefficient functions:
\begin{align}    
   &\Delta = (r - r_-) (r - 2M) + a^2 - (N-N_{-})^2, \nn\\
    &\Sigma = r(r-r_{-}) + (a\cos\theta + N)^2 - N_{-}^2,\nn\\
    &\omega = \frac{-2w}{\Delta - a^2 \sin^2\theta},\quad
     w =  N \Delta \cos\theta
        + a \sin^2\theta \left( M(r-r_{-}) + N(N - N_{-}) \right)\nn.
\end{align}
Two new parameters $r_-,\,N_-$   depend on mass $M$, NUT parameter $N$, electric and magnetic charges $Q,P$ as follows:  
\be r_-=\frac{M(Q^2+P^2)}{M^2+N^2},\qquad N_-=\frac{N(Q^2+P^2)}{2(M^2+N^2)},\ee and $a$ is the rotation parameter.
The electric and magnetic potentials and the complex axidilaton scalar (\ref{zdef}) depend also on the asymptotic values of dilaton and axion     $\z_\infty=\kappa_\infty +i\e^{-2\phi_\infty}$:
\begin{align*}
    &
    v =
    -\frac{\sqrt{2}e^{\phi_\infty}}{\Sigma}
    \text{Re}\left[\mathcal{Q}(r-r_{-} + i\delta)\right],
    \qquad
    u =
    -\frac{\sqrt{2}\e^{\phi_\infty}}{\Sigma}
    \text{Re}\left[
        \mathcal{Q}\z_\infty(r-r_{-} + i\delta)
    \right]
    \\  &
    \z = \frac{
        \z_\infty R + \mathcal{D} \z_\infty^*
    }{
        R + \mathcal{D}
    },\quad
    R = r - \frac{\mathcal{M}^* r_{-} }{ 2M } + i\delta,\quad
    \delta = a \cos\theta + N - N_{-},
\end{align*}
where the   complex mass, electromagnetic and axidilaton charges are introduced
\be
    \mathcal{M} = M + i N,\qquad
    \mathcal{Q} = Q + i P,\qquad
    \mathcal{D} \equiv D+iA= - \frac{{\mathcal{Q}^*}^2}{2\mathcal{M}}.
\ee
 The explicit expressions for the axion and dilaton fields are
\begin{equation}
    \kappa =
     \kappa_\infty - e^{-2\phi_\infty}\frac{2{\rm Im}(R\mathcal{D}^*)}{|R + \mathcal{D}|^2},\qquad
    e^{-2(\phi-\phi_\infty)} = \frac{\Sigma}{|R + \mathcal{D}|^2},
\end{equation}   
where $|R|^2 - |\mathcal{D}|^2 = \Sigma$. The dilaton asymptotics renormalizes the electric and magnetic charges, and the axion determines their mixing. Indeed, for $r\to \infty$, the electric and magnetic potentials are read as follows:
\be
    v\sim -\frac{\sqrt{2}\,Q\,\e^{\phi_\infty} }{r},\qquad
    u\sim \frac{\sqrt{2}\,\left(P\,\e^{-\phi_\infty}-\kappa_\infty Q \e^{\phi_\infty} \right)}{r}.
\ee
The axion and dilaton asymptotics $r\to \infty$ read as follows:
\be
    \z \sim
        \z_\infty
        - 2 i e^{-2\phi_\infty}\frac{\mathcal{D}}{r},\qquad
    \kappa \sim
        \kappa_\infty
        + 2 e^{-2\phi_\infty} \frac{A}{r},\qquad
    \phi \sim \phi_\infty + \frac{D}{r}.
\ee
Combining electric and magnetic potentials, the complex potential $\Psi$ reads
\begin{equation}
    \Psi =
    \frac{\sqrt{2}\e^{\phi_\infty}}{\Sigma}
    \left[
          (\z - \kappa_\infty)
        \text{Re}\left[\mathcal{Q}(r-r_{-} + i\delta)\right]
        + \e^{-2\phi_\infty}\text{Im}\left[
            \mathcal{Q}(r-r_{-} + i\delta)
        \right]
    \right].
\end{equation}

Applying the inverse dualization equations (\ref{udef}),   we find the azimuthal component of the Maxwell four-potential.
For the four-potential one-form ${\rm A}=A_\mu dx^\mu $ we obtain:
\be
 A_\mu dx^\mu =
  \frac{\e^{\phi_\infty}}{\Sigma}
  \left(-Q (r-r_{-}) + P \delta\right)
    \left(
        dt + \left(2Ny - a(1-y^2)\right) d\varphi
    \right) 
    - \e^{\phi_\infty} P y d\varphi.
    \ee
Its norm diverges on the Misner strings $\cos\theta=\pm 1$:
\be
    A_\mu A^\mu \big|_{\cos\theta\to\pm 1}\sim
    \frac{ -\e^{2\phi_\infty}P^2 }{ \Sigma \sin^2\theta },
\ee
thus, Misner rods are also Dirac string loci. Gravitational dualization equation  (\ref{eq:chi}) gives the metric twist potential
\begin{equation}\lb{hi}
    \chi = 2 \frac{
        M\delta - N(r - r_{-})
    }{\Sigma}.
\end{equation}

The ergosphere boundary is given by 
\begin{equation}
    r_E = M +  {r_{-}}/{2} +
  \sqrt{|\mathcal{M}|^2\left(1-r_-/2M\right)^2-a^2\cos^2\theta},
\end{equation}
and the event horizon radius (the outermost root of $\Delta=0$) is equal to
\begin{equation}
    r_H = M +  {r_{-}}/{2} +
  \sqrt{|\mathcal{M}|^2\left(1-r_-/2M\right)^2-a^2},
\end{equation}
so that the horizon two-surface touches the ergosphere at the polar axis.

The transformation to the Weyl coordinates $\rho,\,z$ is given by
\begin{align*}
&\rho^2   =   \Delta   \sin^2\theta,\\
&z  = ( \Delta'/2) \,\cos\theta= (r   -   M-r_-/2) \cos\theta,.
\end{align*}
Solving the equation $\rho=0$, 
we find that the rod system consists of three rods:
\begin{itemize}
    \item  $l_+:$ northern Misner string $r\in(r_H,\infty)$, $\cos\theta = 1$;
    \item $l_H$ the horizon rod $r = r_H$, $\theta\in[0,\pi]$;
    \item $l_+$ southern Misner string $r\in(r_H,\infty)$, $\cos\theta = -1$.
\end{itemize}
The normalized directional vectors
are given by 
the formulas (\ref{lh}), (\ref{ln}) in terms of the corresponding surface gravities and angular velocities. 

It is also useful to represent the solution  in terms of  prolate spheroidal coordinates $x,\,y$ definied as
\begin{equation}
    r=\k x+M+r_-/2,\qquad
    \cos\theta = y,
\end{equation}
\begin{equation} 
    \k^2 = |\mathcal{M}|^2-
    |\mathcal{D}|^2-|\mathcal{Q}|^2 -a^2.
\end{equation}
One obtains:
\begin{align}
     ds^2 =
    - f \left(dt - \omega d\varphi\right)^2
    + f^{-1} \left[
       \k^2 e^{2k}(x^2 - y^2)\left( \frac{dx^2}{x^2-1} + \frac{dy^2}{1-y^2}
        \right)
        + \rho^2 d\varphi^2
    \right],
\end{align}
where the metric functions now read 
\begin{align*}
&\Sigma =
      (\k x + M)^2
    - |\mathcal{D}|^2 
    + (ay + N)^2,\\
    &{\Sigma}f = \k^2 (x^2 - 1) - a^2 (1 - y^2),\quad
    e^{2k} = \frac{f\Sigma}{\k^2(x^2 - y^2)},\\
   &f\Sigma \omega =-  2\left[ N \k^2(x^2-1)y
        + a (1-y^2) \left( M\k x+|\mathcal{M}|^2 -  |\mathcal{Q}|^2/2 \right) \right].
\end{align*}
The South segment of the Misner string is $x>1,\,y=-1$, the horizon rod is $x=1, -1\leq y \leq 1$ and the North Misner string is $x>1,\,y=1$. As expected, the inverse rotation function on the rods is constant on them and represents their angular velocities. It is easy to find
\be
    \Omega_H=  \frac{2a}{ \nu^2},\qquad
    \Omega_{\pm}=\mp \frac1{2N},
\ee 
where a new parameter is introduced
\begin{equation}
    \nu^2 = 2 M \k + 2|\mathcal{M}|^2 - |\mathcal{Q}|^2,
\end{equation}
such that the boundary values of $\Sigma(x,y)$ are
\begin{equation}
    \Sigma_\pm \equiv \Sigma(1,\pm1) = \nu^2 \pm 2 a N.
\end{equation}
The other ingredients of the mass formulas depend on the values of electric and magnetic potentials $v_\pm$ and $u_\pm$ of the defect rods  $l_{\pm}$. 
The magnetic potential in spheroidal coordinates is
\begin{equation}
    u = \frac{\sqrt{2}}{\Sigma}\Bigr\{
          \e^{\phi_\infty} \kappa_\infty [-Q (\k x + M - r_-/2) + P \delta]
        + e^{- \phi_\infty } [P (\k x + M - r_-/2) + Q \delta]
    \Bigr\}.
\end{equation}
Using it, we obtain:  
\begin{align}\lb{vpm}
    &
    v_{\pm} =
    -\frac{\sqrt{2}e^{\phi_\infty}}{\nu^2 \pm 2 a N}
    \text{Re}\left[\mathcal{Q}(
          \k \pm i a
        + \mathcal{M}
        (1 - |\mathcal{Q}|^2/2|\mathcal{M}|^2)
    )\right],
    \\\lb{upm} &
    u_{\pm} =
    -\frac{\sqrt{2}\e^{\phi_\infty}}{\nu^2 \pm 2 a N}
    \text{Re}\left[
        \mathcal{Q}\z(
          \k \pm i a
        + \mathcal{M}
        (1 - |\mathcal{Q}|^2/2|\mathcal{M}|^2)
    )
    \right].
\end{align}

For surface gravities it is convenient to use Eq. (\ref{kappa_weyl}):
\begin{equation}
    \varkappa_H= 
    \frac{\k}{\nu^2} ,\qquad
    \varkappa_\pm=\mp\frac{1}{2N}.
\end{equation}angular velocities and surface gravities of Misner strings coincide similarly to the case of Kerr-Newman-NUT \cite{Clement:2019ghi}. But they enter into the mass formula in a different way, so one should not consider this coincidence as a matter of interpretation.  

The electric charges of the rods are given by (\ref{Qn}).
One finds the Gauss charges of the horizon and  strings as: 
\begin{align}
    Q_H = \frac{\omega_H}{2\sqrt{2}} (u_+ - u_-) =
    \frac{
        e^{-\phi \infty }
        \nu^2
        \left(2 \k N^2 + M\nu^2\right)
        \text{Im}\left[\z_\infty e^{2 \phi_\infty } \mathcal{M} \mathcal{Q}\right]
    }{
        |\mathcal{M}|^2
        \left(\nu^4 - 4 a^2 N^2\right)
    },
\end{align}
\begin{equation}
Q_\pm = \frac{N u_\pm}{\sqrt{2}}.
\end{equation}
Combining them together, we obtain
\begin{equation}
    Q_H + Q_{+} + Q_{-} = Q e^{-\phi_\infty } + P \kappa_\infty e^{\phi_\infty }.
\end{equation}
This coincides with the true asymptotic charge of the configuration, with account of axion mixing and dilaton renormalization. 

The electric potential on the horizon and on the strings are
\begin{equation}
    \Phi_H = 
    e^{\phi \infty } 
    \frac{
          \nu^2 (Q M - N P) + 2\k N (Q N + M P)
    }{2 \nu^2 |\mathcal{M}|^2},\qquad
    \Phi_\pm = -e^{\phi_\infty}\frac{P}{2 N}.
\end{equation}

Rod masses can be expressed in the form (\ref{Mrn1}), using the corresponding values of the twist potential (\ref{hi}). One obtains
\begin{align}
    &
    M_H =
    \nu^2\cdot
    \frac{
        M \nu^2 + 2\k N^2
    }{\nu^4 - 4a^2N^2}
    +
    \frac{1}{4}
    (
        2N (v_{+} u_+ + v_{-} u_-)
        - \sqrt{2} \e^{\phi_\infty} P(u_+ + u_-)
    ),
\end{align}
\begin{equation}
    M_\pm =
    N \frac{
        - N \k \pm M a
    }{\nu^2 \pm 2 a N}
    +\frac{1}{4} (2N v_\pm - \sqrt{2} P \e^{\phi_\infty} ) u_\pm,
\end{equation}
where the boundary values of electric and magnetic potentials are given by (\ref{vpm}), (\ref{upm}).
One can distinguish Dirac string contributions to the masses of defects as the parts non-vanishing for $N=0$. Other terms are due to Misner strings. Direct substitution of (\ref{vpm}-\ref{upm}) leads to long expressions, but one can show that the Maxwell contributions to the total mass cancel out, while the remaining part non-trivially leads to the identity
$$M_{\infty} = M_H + M_{+} + M_{-} = M.$$

Now we turn to the angular momentum balance, using the rod's momenta the expression (\ref{Jrn1}). The length of the horizon rod  is $L_H = 2\k$. Other rod lengths diverge. As it was shown in Ref. \cite{Clement:2019ghi}, to overcome this issue, one can regularize the angular momentum as follows:
\begin{equation}
    \tilde{J}_\pm = J_\pm + \frac{\omega_\pm L_\pm}{4}.
\end{equation}
We then obtain the angular momenta of the three rods as
 \begin{align}
  J_H = 
      \frac{\omega_H}{2}\left(
        -\k + M_H - Q_H \Phi_H
      \right),\qquad
  \tilde{J}_{\pm} = 
    N^2 \left(
        \frac{
            \pm N \k - M a
        }{\nu^2 \pm 2 a N}
        \mp \frac{v_\pm u_\pm}{2}
    \right).
 \end{align}
We have explicitly checked that
$$\tilde{J}_{\infty} = J_H + \tilde{J}_+ + \tilde{J}_- = a M.$$

Therefore, we have constructed horizon mass, angular momentum and charge in terms of the global parameters and checked their total balance.  
 
\section{Conclusion}
We have shown that the derivation of mass formulas using the Tomimatsu approach for calculating Komar integrals around rods in electrovacuum  can be generalized to the EMDA theory containing scalar fields coupled to the graviphoton. Using a three-dimensional reduction of the EMDA equations, we
construct magnetic and twist potentials, which allow to express the integrals over the rods   in the same way as in the Einstein-Maxwell theory, where Tomimatsu used for this purpose the Ernst potentials. 
It should be noted that the scalar dilaton and axion fields do not contribute to the resulting mass formulas, and their asymptotics only rotates electric and magnetic charges in the parameter space. For regular asymptotically flat configurations, the dilaton and axion charges are secondary  and their variations do not enter the Smarr-type formulas.   

In this approach, the magnetic and NUT charges do not affect the mass and angular momentum of the black hole calculated as Komar's integrals over the horizon. They contribute to the asymptotic mass and angular momentum via the respective individual rods representing the Dirac and Misner strings. As an illustration, we considered the case of a rotating EMDA dyon equipped with NUT. In this case, as in the case of dyonic Kerr-Neumann-NUT, the Dirac string contributes to the asymptotic mass, but not to the asymptotic angular momentum, while the Misner string contributes to both. 

The work is supported by the Russian Foundation for Basic Research on the project 20-52-18012Bulg-a, and the Scientific and Educational School of Moscow State University “Fundamental and Applied Space Research”. I.B. is also grateful to the Foundation for the Advancement
of Theoretical Physics and Mathematics “BASIS” for support.

\end{document}